\documentclass[12pt,a4paper]{article}
\begin{document}
\begin{center}
{\bf
Do neutrino oscillations allow an extra phenomenological
parameter?} \\[3mm]
I.S. Tsukerman \\[2mm] 
State Research Center  \\
``Institute for Theoretical and 
Experimental Physics'' \\
Moscow, 117218, Russia  \\
E-mail:zuckerma@heron.itep.ru \\
\vspace*{5mm}

{\it Submitted to JETP Letters}
\end{center}

\vspace*{5mm}

\begin{abstract}

The quantity $\xi$ introduced recently in the phenomenological
description of neutrino oscillations is in fact not a free
parameter, but a fixed number.

\end{abstract}

The literature on phenomenology of neutrino oscillations is vast
(see, e.g., \cite{1} - \cite{6} and references therein). In a
recent paper \cite{7} Giunti and Kim in the case of two-flavour
mixing have introduced a new phenomenological parameter $\xi$.
According to \cite{7}, $\xi = 0$ corresponds to the so-called
equal momentum assumption \cite{1,2}, while $\xi =1$ corresponds
to equal energy assumption \cite{5,6}. Authors of \cite{7}
emphasize that $\xi$ disappears from final expressions for the
neutrino oscillation probability.

The aim of this note is to indicate that parameter $\xi$ is fixed
by energy-momentum conservation in the process which is
responsible for neutrino emission, as explicitly assumed in ref.
\cite{7}.

Following ref. \cite{7} we will consider the decay $\pi \to
\mu\nu$ in the framework of two-flavour toy model.  The parameter
 $\xi$ is defined in \cite{7} for the pion rest-frame by considering
the auxiliary case of absolutely massless neutrinos and denoting
the energy of such neutrinos as $E$,
\begin{equation}
\xi = ~^1/_2(1+ m_{\mu}^2/m_{\pi}^2) \;\; , \label{5}
\end{equation}
where $m_{\mu}$ and $m_{\pi}$ are the masses of the muon and the
pion. Then for massive (but light!) neutrinos they get:
\begin{equation}
E_{1,2} = E+(1-\xi)m_{1,2}^2/2E \;\; , \label{1}
\end{equation}
\begin{equation}
p_{1,2} = E-\xi m_{1,2}^2/2E \;\; . \label{2}
\end{equation}
Here $E_{1,2}$, $p_{1,2}$ and $m_{1,2}$ are the energies, momenta
and masses of the neutrinos, respectively. Wherefrom the above
statement about $\xi = 0, 1$ follows:
\begin{equation}
E_1 = E_2 \;\; {\rm for} \;\; \xi =1 \;\; {\rm and} \;\; p_1 = p_2 \;\; {\rm for} \;\; \xi =0 \;\;.
\label{6}
\end{equation}
Thus, the equal energy and equal momentum assumptions in the form
$\Delta E \equiv E_1 -E_2 =0$ and $\Delta p \equiv p_1 - p_2 =0$,
respectively, are treated by authors of ref. \cite{7} as
particular cases of the general kinematical relations  (\ref{1})
and (\ref{2}):
\begin{equation}
\Delta E = (1-\xi)\Delta m^2/2E = 0 \;\; {\rm for} \;\; \xi =1
\;\;, \label{7}
\end{equation}
\begin{equation}
\Delta p = \xi\Delta m^2/2E = 0 \;\; {\rm for} \;\; \xi =0 \;\; .
\label{8}
\end{equation}

Unfortunately, both the treatment and the relations (\ref{6}) -
(\ref{8}) are erroneous.

On  one hand, the quantity $\xi$ is not a free parameter for a
certain decay process. Indeed, it follows from (\ref{5}) that
$\xi$ has a fixed value ($\simeq 0.8$) for the decay under
consideration. On the other hand, it is evident from definitions
of $E$ and $\xi$ that
\begin{equation}
E = m_{\pi}(1-\xi) \;\; . \label{3}
\end{equation}
The parameter $\xi$ determines sharing of the decay energy. As
seen from (\ref{3}), the values $\xi =0$ and $\xi =1$ are
senseless ones because they refer accordingly to the limiting
cases of $E_{\mu} = 0$ and $E = 0$.  Therefore  one cannot assume
that $\xi$ can be equal to 1 or 0. Instead of that, the solution
of the equalities (\ref{7}) and (\ref{8}) is the vanishing $\Delta
m^2$, that is absence of the oscillations. \\


{\bf \large Acknowledgements}

\vspace{3mm} The author is grateful to L.B. Okun for his friendly
support. This work was supported by RFBR grant 00-15-96562.

\end{document}